\documentclass[fleqn,twoside]{article}
\usepackage{espcrc2}

\usepackage{graphicx}
\usepackage[figuresright]{rotating}

\newcommand{\AmS}{{\protect\the\textfont2
  A\kern-.1667em\lower.5ex\hbox{M}\kern-.125emS}}

\def\apj{ApJ}%
\def\apjl{ApJ}%
\def\aap{A\&A}%
\def\pasj{PASJ}%
\def\nat{Nature}%

\def \nineteen {4U\,1916-053}
\def \mxb {MXB\,1658-298}
\def \bigdip {X\,1624-490}
\def \twelve {X\,1254-690}
\def \grs {GRS\,1915+105}
\def \gro {GRO\,J1655-40}
\def \gx {GX\,13+1}

\def\countsec{\hbox{counts s$^{-1}$}}

\def\degmark{^\circ}
\newcommand {\degree} {$^{\circ}$}

\newcommand {\eline} {$E_{\rm line}$}

\newcommand {\sig} {$\sigma$}

\newcommand {\fetfive} {Fe\,{\sc xxv}}
\newcommand {\fetsix} {Fe\,{\sc xxvi}}
\newcommand {\nitseven} {Ni\,{\sc xxvii}}
\newcommand {\ssixteen} {S\,{\sc xvi}}

\newcommand {\ka} {K$\alpha$}
\newcommand {\kb} {K$\beta$}

\title{Discovery of X-ray absorption lines from the low-mass X-ray binaries \nineteen\ and \twelve\ with XMM-Newton}

\author{L. Boirin\address[ESTEC]{Astrophysics Missions Division, 
		Research and Scientific Support
		Department of ESA, \\ ESTEC, Postbus 299, NL-2200 AG
		Noordwijk, The Netherlands}%
        	\thanks{L. Boirin acknowledges an ESA Fellowship.},
	A.~N. Parmar\addressmark[ESTEC],
	D. Barret\address[CESR]{Centre d'Etude Spatiale des Rayonnements, 
		CNRS/UPS,\\ 9 Av. du Colonel Roche, 31028 Toulouse Cedex 4, 
		France},
	S. Paltani\address[Marseille]{Laboratoire d'Astrophysique de Marseille,
		Traverse du Siphon, BP 8, 13376 Marseille Cedex 12, France}}
       
\begin{document}

\begin{abstract}
We report the discovery of narrow X-ray absorption features from the
two dipping low-mass X-ray binary \nineteen\ and \twelve\ during
XMM-Newton observations. The features detected are identified with
resonant scattering absorption lines of highly ionized iron (Fe XXV
and Fe XXVI).  Resonant absorption features are now observed in a
growing number of low-mass X-ray binaries (LMXBs): the two
superluminal jet sources \grs\ and \gro, the bright LMXB \gx\ and the
four dipping sources \mxb, \bigdip, \nineteen\ and \twelve.  The early
hypothesis that their origin could be related to the presence of
superluminal jets is thus ruled out. Ionized absorption features may
be common characteristics of accreting systems.  Furthermore, their
presence may depend on viewing angle, as suggested by their detection
in dippers which are viewed close to the disk plane, and by the fact
that \grs, \gro\ and \gx, although not dippers, are suspected to be
also viewed at high inclination.
\vspace{1pc}
\end{abstract}

\maketitle

\section{INTRODUCTION}

\subsection{\nineteen\ and \twelve}

\nineteen\ and \twelve\ are two LMXBs showing periodic intensity dips
in their X-ray light curves. The dips recur at the orbital period of
the system. Dips are believed to be due to obscuration of the central
X-ray source by vertical structure located at the outer edge of the
accretion disk and due to the impact of the accretion flow from the
companion star into the disk \cite{1916:white82apjl}.  The presence of
dips in these two sources and the lack of X-ray eclipses from the
companion star indicate that the system is viewed relatively close to
edge-on, at an inclination angle in the range $\sim$60--80\degree\
\cite{frank87aa}.  The period of the dips is 50 minutes in \nineteen\
\cite{1916:white82apjl,1916:walter82apjl} and 3.88~h in \twelve\
\cite{1254:courvoisier86apj}.

\nineteen\ and \twelve\ have been observed using a variety of X-ray
instruments. In \twelve\, the dips were not detected during all
observations, but the dipping activity ceased and re-appeared several
times. A likely explanation for the absence of dips during some
observations is that the vertical structure in the outer region of the
accretion disk had decreased in size, so that the central X-ray source
could be viewed directly \cite{1254:smale99apj}.

\subsection{Narrow absorption features in X-ray binaries}

Narrow absorption features from highly ionized Fe and other metals in
the spectra of LMXBs were first seen from \gx\ using ASCA by Ueda et
al.  \cite{gx13:ueda01apjl} who detected a narrow absorption feature
at 7.01~keV which they interpreted as resonant scattering of the
K$\alpha$ line from H-like Fe.  XMM-Newton observations revealed an
even more complex picture for \gx\ with narrow absorption features
identified with the K$\alpha$ and K$\beta$ transitions of He- and
H-like iron (Fe\,{\sc xxv} and Fe\,{\sc xxvi}) and H-like calcium
(Ca\,{\sc xx}) K$\alpha$ detected \cite{gx13:sidoli02aa}.  There is
also evidence for the presence of a deep Fe~{\sc xxv} absorption edge
at 8.83~keV and of a broad emission feature around 6.4~keV.
XMM-Newton observations of the eclipsing and dipping LMXB \mxb\
revealed narrow absorption features identified with O\,{\sc
viii}~K$\alpha$, K$\beta$, and K$\gamma$, Ne\,{\sc x} K$\alpha$,
Fe\,{\sc xxv}~K$\alpha$, and Fe\,{\sc xxvi}~K$\alpha$ together with a
broad Fe emission feature at 6.47~keV \cite{1658:sidoli01aa}.  Another
LMXB dip source, \bigdip, also displays K$\alpha$ absorption lines
identified with Fe\,{\sc xxv} and Fe\,{\sc xxvi}, as well as fainter
absorption features tentatively identified with Ni\,{\sc xxvii}
K$\alpha$ and Fe\,{\sc xxvi} K$\beta$.  A broad emission feature at
6.58~keV is also evident \cite{1624:parmar02aa}.

The properties of the absorption features in the 3 LMXBs show no
obvious dependence on orbital phase, except during a dip from \bigdip,
where there is evidence for the presence of additional cooler
material.  Previously, the only X-ray binaries known to exhibit such
narrow X-ray absorption lines were two superluminal-jet sources, and
it had been suggested that these features are related to the jet
formation mechanism. This now appears unlikely, and instead, as we
discuss below, their presence is probably related to the viewing angle
of the system.

Here we report the discovery of narrow X-ray absorption features from
highly ionized Fe in the XMM-Newton spectra of \nineteen\ and \twelve.
Detailed reports on these discoveries may be found in
\cite{1916:boirin03aa} and \cite{1254:boirin03aa} respectively.

\section{OBSERVATIONS \& RESULTS}

\nineteen\ was observed by XMM-Newton \cite{jansen01aa} for 5~h on
2002 September 25. \twelve\ was observed for 4~h on 2001 January 25,
and for 7~h one year later on 2002 February 7. We report results
obtained with the EPIC pn camera \cite{turner01aa} in the 0.5--10~keV
energy band.

\subsection{X-ray light curves}

\begin{figure}[!h]
\centerline{\hspace{-0.3cm}\includegraphics[angle=90,width=0.53\textwidth]{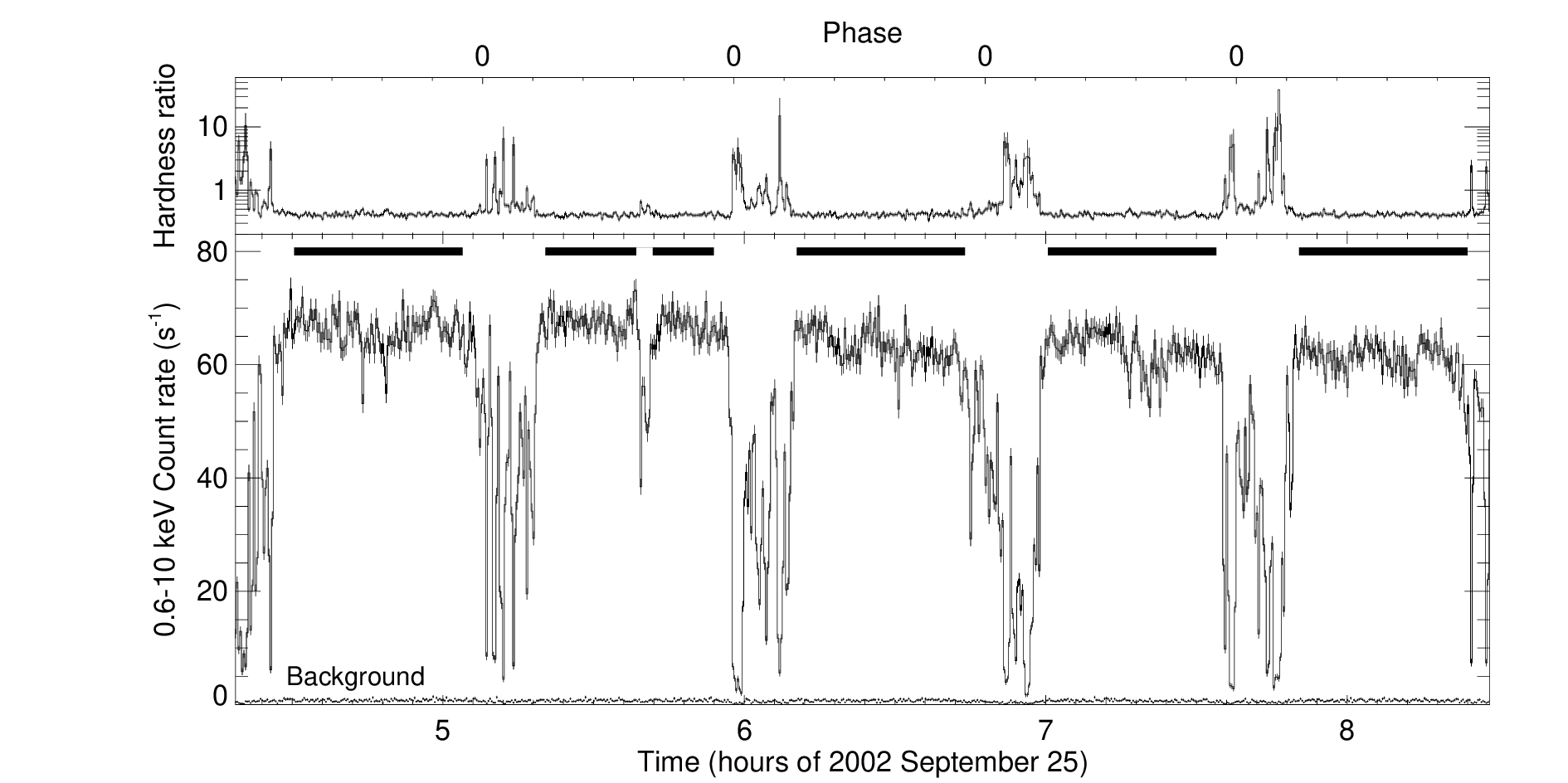}}
\vspace{-0.3cm}
\centerline{\hspace{-0.3cm}\includegraphics[angle=90,width=0.53\textwidth]{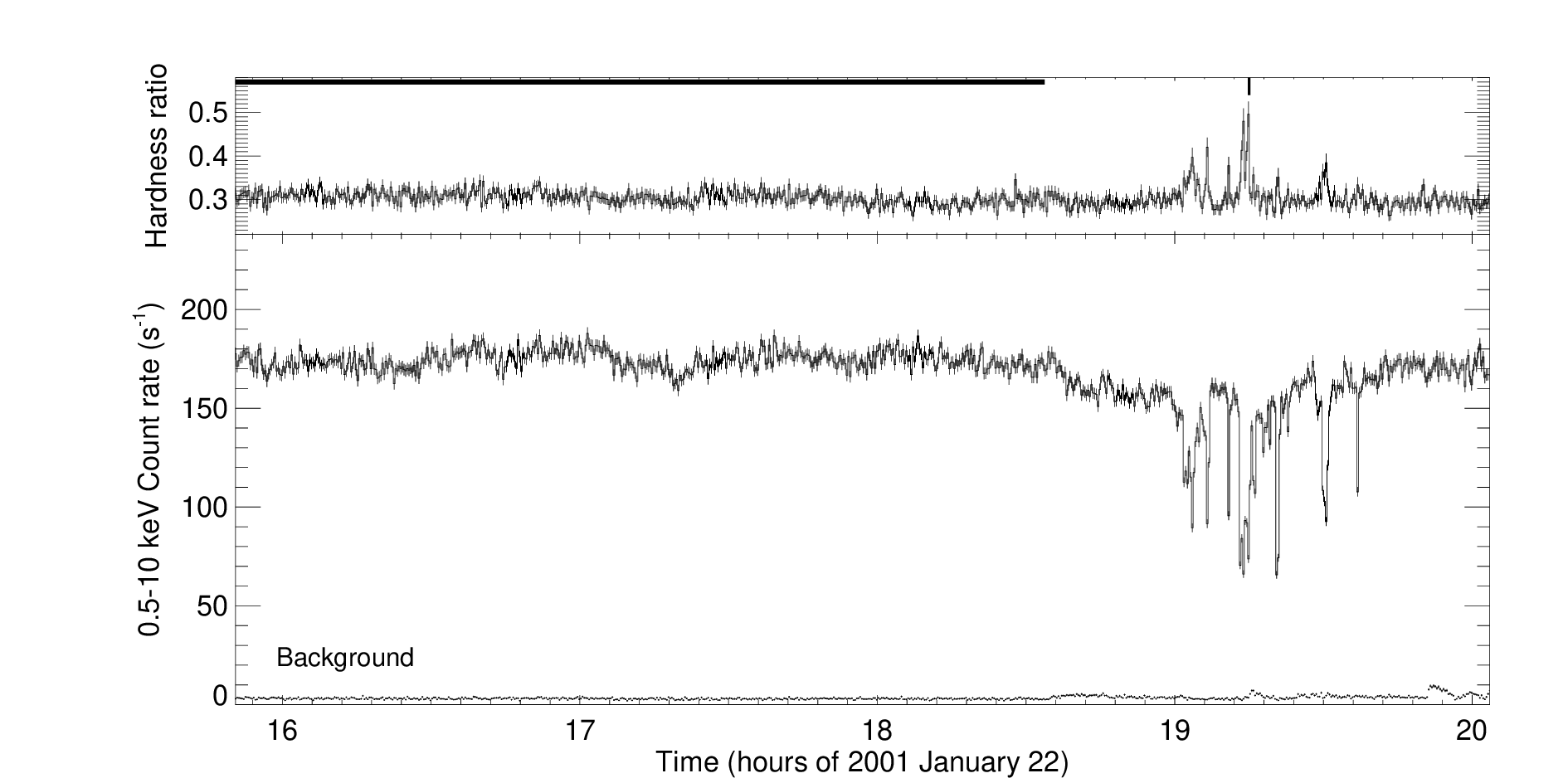}}
\vspace{-0.3cm}
\centerline{\hspace{-0.3cm}\includegraphics[angle=90,width=0.53\textwidth]{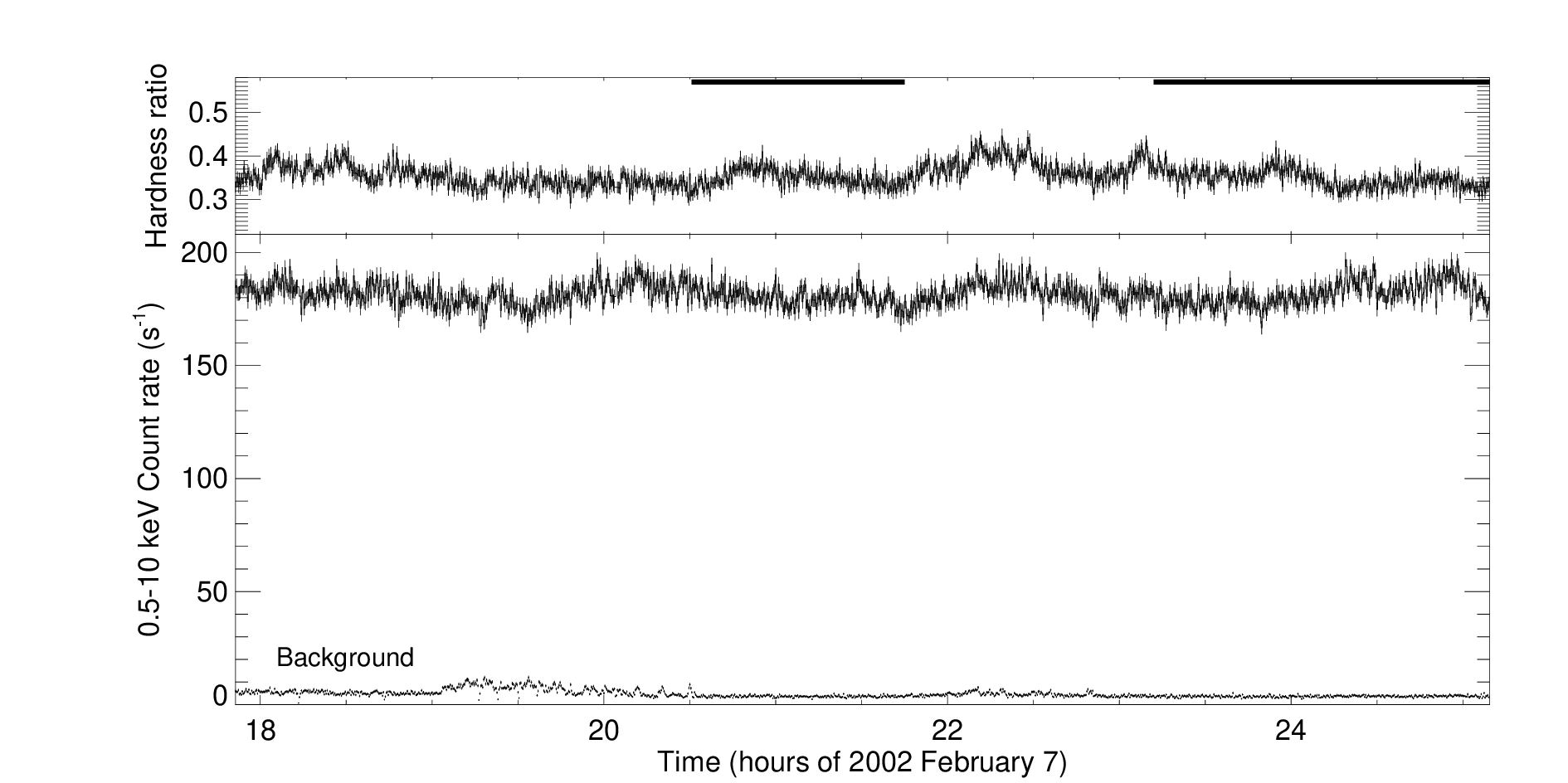}}
\vspace{-1cm}
\caption{The 0.5--10~keV EPIC pn background subtracted light curves of
\nineteen\ on 2002 September (top), \twelve\ on 2001 January (middle),
and \twelve\ again on 2002 February (bottom).  The upper panels show
the hardness ratio (counts in the 2.5--10~keV band divided by counts
in the 0.5--2.5~keV band). The binning time is 20~s in every panel.}
\label{fig_pn_lc}
\end{figure}

Fig.~\ref{fig_pn_lc} shows the EPIC pn light curves of \nineteen\ on
2002 September (top), \twelve\ on 2001 January (middle), and \twelve\
again on 2002 February (bottom).  The upper panels show the hardness
ratio.  The periodic dipping activity of \nineteen\ is clearly visible
and associated with spectral hardening. The dipping intensity shows
erratic variability. The persistent intensity (outside the dips)
decreases slowly during the observation from $\sim$67 to
$\sim$62~\countsec.  A dip, also associated with spectral hardening,
is clearly visible during the 2002 observation of \twelve. Remarkably,
there is no evidence for any dipping activity from \twelve\ one year
later during the 2002 observation, which covers nearly two orbital
periods of the system. Such a cessation and re-appearance of the
dipping activity has already been observed
\cite{1254:smale99apj,1254:iaria01apj,1254:smale02apj}.  The
persistent intensity of \twelve\ is $\sim$185~\countsec\ and
$\sim$175~\countsec\ during the 2001 and 2002 observations,
respectively.

\subsection{X-ray spectra}

\subsubsection{Persistent emission}

\begin{figure*}[!t]
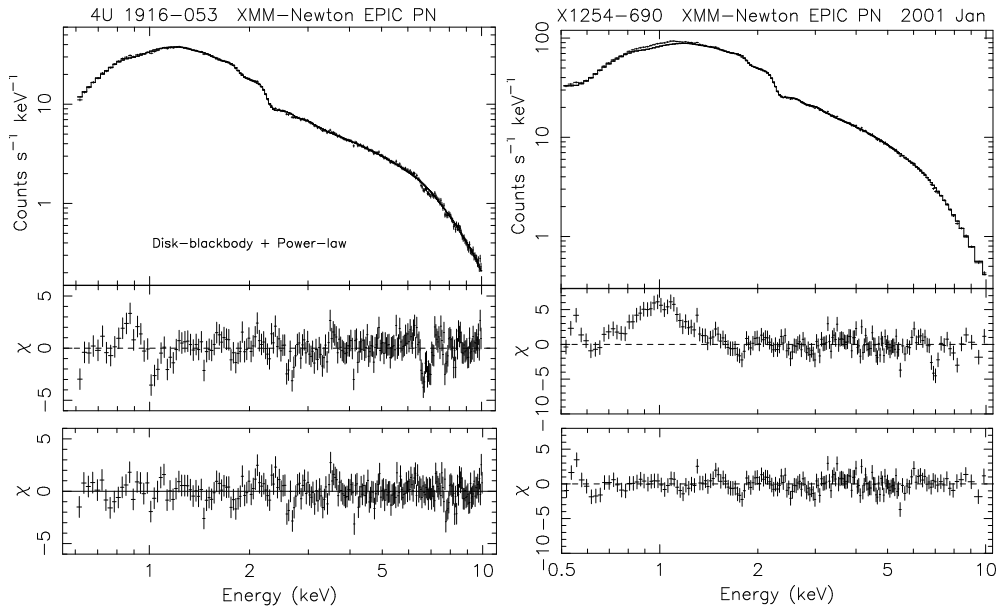

\centerline{
  \hbox{\hspace{0.0cm}
   \includegraphics[height=6.5cm,angle=-90]{fig_boirin_1915_spectrum.ps}
   \includegraphics[height=6.5cm,angle=-90]{fig_boirin_1254_spectrum_2001.ps}}
  \vspace{-0.1cm}
}
\centerline{
  \hbox{\hspace{0.15cm}
   \includegraphics[height=6.4cm,angle=-90]{fig_boirin_1915_res_with_gauss.ps}
   \hspace{0.0cm}
   \includegraphics[height=6.4cm,angle=-90]{fig_boirin_1254_res_with_gauss_2001.ps}}
}

  \caption[]{The upper panels show the EPIC pn spectra of the
             persistent emission of \nineteen\ (left) and of \twelve\
             during the 2001 observation, and the best-fit
             disk-blackbody and power-law continuum models. The
             residuals (middle panels) reveal the presence of a broad
             feature centered around 1~keV, together with several
             narrow absorption features around 7~keV.  The lower
             panels show the residuals when these features are
             included in the spectral model.}  
\label{fig:spectrum}
\end{figure*}

Persistent-emission spectra were extracted from the \nineteen, the
\twelve\ 2001 and 2002 observations by selecting non-dipping
intervals.  The overall continua were modeled using absorbed
multicolor disk-blackbody and power-law components. The spectra fit
with such models are shown in Fig.~\ref{fig:spectrum} (upper
panels). The spectrum of \twelve\ during the 2002 observation is not
shown, as it is very similar to the 2001 observation one.

Examination of the residuals (middle panels of
Fig.~\ref{fig:spectrum}) reveals broad structure around 1~keV. This
structure is well modeled by an edge at $0.98 \pm 0.02$~keV with an
optical depth, $\tau$, of $0.10 \pm 0.02$ in the case of \nineteen, by
a Gaussian with an energy of $0.96 \, ^{+0.04} _{-0.06}$~keV, a width
($\sigma$) of $175 \, ^{+75} _{-50}$~eV in the case of the \twelve\ in
2001, and by a similar Gaussian in the case of \twelve\ in 2002. In
the case of \twelve, this is probably the same structure noted by
Iaria et al. \cite{1254:iaria01apj} using BeppoSAX and modeled as an
absorption edge at 1.27~keV with $\tau$ of 0.15. The nature of the
$\sim$1~keV structure in \nineteen\ and in \twelve\ is unclear. The
smaller structures at $\sim$1.8~keV and $\sim$2.2~keV are probably due
to an incorrect instrumental modeling of the Si CCD and Au mirror
edges.

\begin{table}[!t]
\begin{center}
\scriptsize
\caption[]{Properties of the absorption Gaussian lines detected in the
EPIC pn spectra of the persistent emission of \nineteen\ and
\twelve. The underlying continuum models include a disk-blackbody, a
power-law and an edge or a Gaussian emission feature near 1~keV. }
\begin{tabular}{llccc}
\hline
\hline
\noalign {\smallskip}
 \multicolumn{5}{c}{Persistent Emission} \\
\noalign {\smallskip}
\hline
\noalign {\smallskip}
&  & {\nineteen}           &  \multicolumn{2}{c}{\twelve} \\
  &  & & 2001 Jan.             &  2002 Feb. \\
\noalign {\smallskip}
\hline
\noalign {\smallskip}

\multicolumn{5}{l}{\fetfive\ \ka\ absorption Gaussian}\\
& \eline\ (keV) & 6.65 $^{+0.05}_{-0.02}$ & &\\
& \sig\ (eV)	& $<$100	& & \\
& $EW$ (eV)		& $-30~^{+8}_{-12}$ & & \\

\noalign {\smallskip}

\multicolumn{5}{l}{\fetsix\ \ka\ absorption Gaussian}\\
& \eline\ (keV) & 6.95 $^{+0.05}_{-0.04}$ &$6.95 \pm 0.03$    & $6.96 \pm 0.04$ \\
& \sig\ (eV)	& $<$140 	 &$<$120  &  $<$95	\\
& $EW$ (eV)	& $-30~^{+11}_{-12}$  &$-27 \, ^{+11} _{-8}$  
& $-21 \, ^{+8} _{-5}$    \\

\noalign {\smallskip}
\multicolumn{5}{l}{\fetsix\ \kb\ absorption Gaussian}\\
& \eline\ (keV)      & &$8.20 \, ^{+0.05} _{-0.10}$  & $8.16 \pm 0.06$ \\
& \sig\ (eV)          & & $<$170     &   $<$80    \\
& $EW$              (eV)         & & $-17 \pm 9$ & $-16 \pm 9$ \\

\noalign {\smallskip}                       
\hline
\label{tab:lines}
\end{tabular}
\end{center}
\end{table}

\begin{table}[!t]
\begin{center}
\scriptsize
\caption[]{Properties of the \fetsix\ \ka\ absorption Gaussian line
detected in 3 individual segments of persistent emission during the
2001 observation of \twelve.}
\begin{tabular}{llccc}
\hline
\hline
\noalign {\smallskip}
\multicolumn{5}{c}{\twelve\ Persistent emission (2001 Jan.)}\\
\noalign {\smallskip}
\hline
\noalign {\smallskip}
 & & Segment 1 & Segment 2 & Segment 3 \\
\multicolumn{2}{l}{Phase range} & 0.12-0.35 & 0.35-0.59 & 0.59-0.82 \\
\noalign {\smallskip}
\hline
\noalign {\smallskip}
\multicolumn{5}{l}{\fetsix\ \ka\ absorption Gaussian}\\
&\eline\ (keV)     &$6.93 \pm 0.09$    & $6.95 \, ^{+0.03} _{-0.06}$ & $6.97 \, ^{+0.04} _{-0.05}$ \\
&\sig\ (eV)                  &$<$185  &    $<$140  & $<$126 \\
&$EW$              (eV)         &$-22 \, ^{+16} _{-15}$  &  $-21 \pm 11$ & $-31 \, ^{+14} _{-12}$ \\

\hline
\label{tab:orbital}
\end{tabular}
\end{center}
\end{table}

\begin{figure*}[!t]
\centerline{\includegraphics[angle=0,height=0.25\textheight]{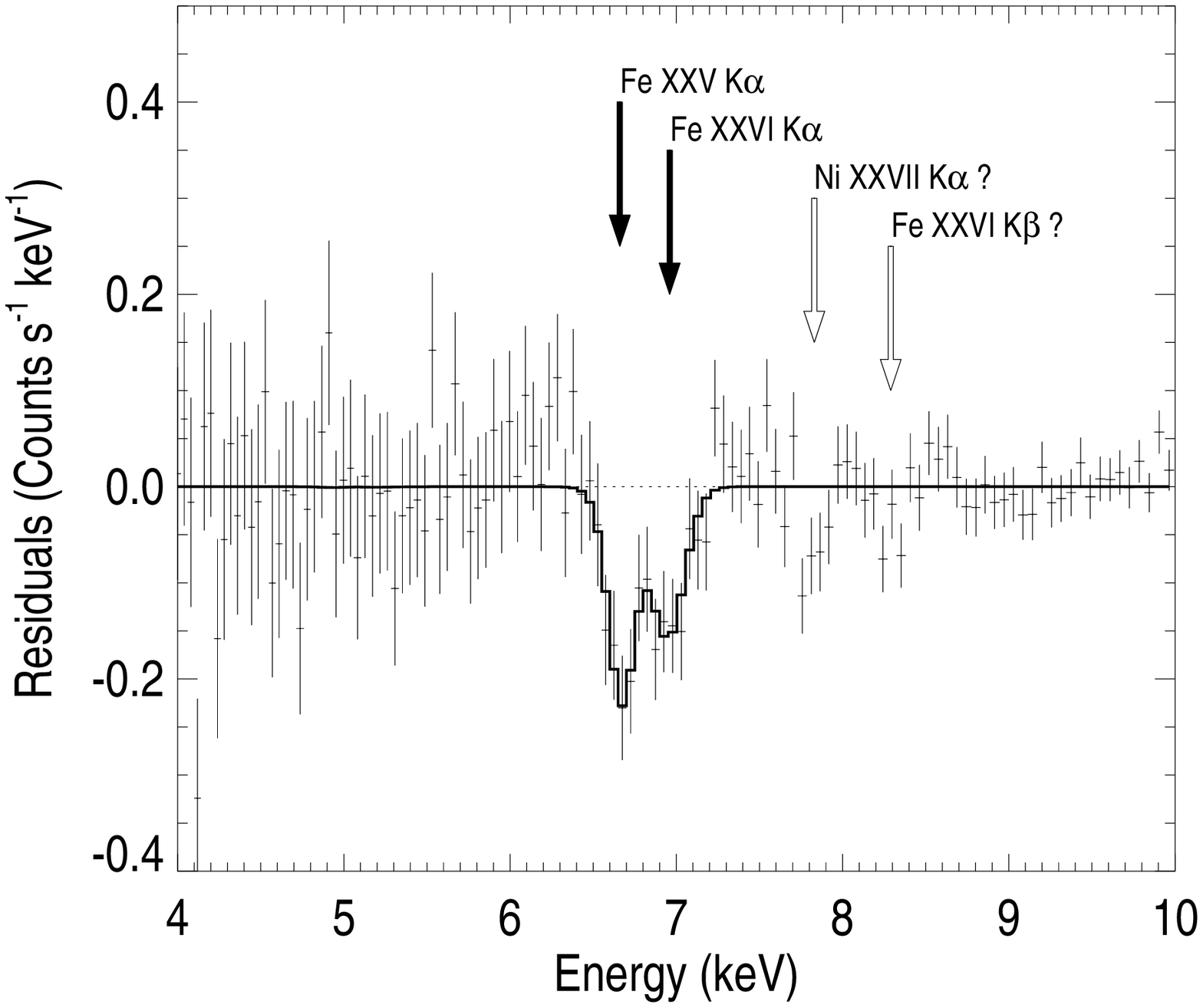}
\includegraphics[angle=-90,height=0.25\textheight,origin=c]{fig_boirin_1254_res_zoom_2001.ps}}
\caption{Zoom of the residuals around 7~keV for \nineteen\ (left) and
\twelve\ (right) when the best-fit models discussed in the text are
fit to the pn spectra of the persistent emission and the
normalisations of the narrow absorption features are set to zero.}
\label{fig:res_zoom}
\end{figure*}

Examination of the remaining fit residuals shows several deep negative
residuals around 7~keV (Fig.~\ref{fig:res_zoom}). These were modeled
by Gaussian absorption lines whose properties (the line energy,
\eline, the width, \sig, and the equivalent width, $EW$) are given in
Table~\ref{tab:lines}, and whose significance was checked to be higher
than 3\sig\ using F-tests. This is the first detection of such
absorption features in \nineteen\ and \twelve. The measured energy of
the Gaussian lines are consistent with \fetfive\ \ka\ and \fetsix\
\ka\ transitions in the case of \nineteen, and \fetsix\ \ka\ and \kb\
transitions in the case of \twelve.  In addition to these clearly
detected lines, in the case of \nineteen, there is marginal evidence
for three other absorption features at 7.82~keV, 8.29~keV, and
2.67~keV at energies consistent with \nitseven\ \ka, \fetsix\ \kb\ and
\ssixteen\ transitions, respectively.

In order to investigate whether the properties of the absorption
features depend on orbital phase, the persistent emission interval of
the 2001 observation of \twelve\ was divided into three intervals of
$\sim$3220~s.  To estimate the phases covered by these intervals, we
use a period of 3.88~h and the time 19.25~h of 2001 January 22,
corresponding to the apparent dip center, as an arbitrary reference
for phase 0.  The same continuum model as the one used for the total
persistent spectrum was fit to the 3 individual spectra. The \fetsix\
\ka\ absorption feature is clearly evident, but the \fetsix\ \kb\
absorption feature is not significantly detected. The properties of
the \fetsix\ \ka\ absorption features in the 3 segments are given in
Table~\ref{tab:orbital}.  They are all consistent with those obtained
from the total persistent spectrum. Thus, there does not appear to be
any obvious dependence of the \fetsix\ \ka\ absorption feature
properties on orbital phase during the persistent emission.

\subsubsection{Dipping emission}

In order to investigate the properties of eventual absorption features
during the dipping intervals, three spectra were extracted during the
dipping emission of \nineteen, based on intensity selection criteria.
Events corresponding to a background-subtracted pn count rate in the
range 40--60, 20--40 and 0--20~s$^{-1}$ (see Fig.~\ref{fig_pn_lc})
were extracted to form ``shallow'', ``intermediate'' and ``deep''
dipping spectra, respectively.  The same continuum model as the one
used for the total persistent spectrum was fit to the 3 individual
dipping spectra with all the parameters fixed except the
normalizations and the absorbing column density, which were allowed to
vary in order to account for the spectral variations of the continuum
during dipping.  This models the overall continuum shape reasonably
well. Furthermore, an absorption feature at $\sim$6.7~keV is also
detected in each intensity-selected spectrum. Its energy is consistent
with a \fetfive\ \ka\ transition, as in the persistent spectrum. The
deep dipping spectrum also shows a second absorption feature at
$\sim$8.35~keV. Its energy is roughly consistent with a \fetsix\ \kb\
transition. However, this identification is inconsistent with the
absence of the \fetsix\ \ka\ line in the spectrum. Thus, the
interpretation of the feature at 8.35~keV is unclear.

\begin{table}[!t]
\begin{center}
\scriptsize
\caption[]{Properties of the absorption Gaussian lines detected in the
three intensity-selected dipping spectra (shallow, intermediate and
deep) of \nineteen.}
\begin{tabular}{llccc}
\hline
\hline
\noalign {\smallskip}
 \multicolumn{5}{c}{\nineteen\ Dipping emission}\\
\noalign {\smallskip}
\hline
\noalign {\smallskip}
 & & Shallow & Inter. & Deep \\
\noalign {\smallskip}
\hline
\noalign {\smallskip}
\multicolumn{5}{l}{\fetfive\ \ka\ absorption Gaussian}\\
&\eline\ (keV) &  6.70 $\pm$ 0.05 & 6.67 $^{+0.06}_{-0.05}$ & 6.59 $\pm$ 0.05\\
&\sig (eV) & $<$110 & 160 $^{+80}_{-60}$ & $<$173\\
&$EW$ (eV) & -74 $^{+20}_{-27}$ & -168 $^{+44}_{-46}$ & -119 $^{+45}_{-50}$\\
\noalign {\smallskip}

\multicolumn{5}{l}{Second absorption Gaussian}\\
&\eline\ (keV) & & & 8.35 $\pm$ 0.05\\
& \sig\ (eV) & & & $<$144 \\
& $EW$ (eV) & & & -120 $\pm$ 50\\
\noalign {\smallskip}
\hline
\label{tab:dipping}
\end{tabular}
\end{center}
\end{table}

\section{DISCUSSION}

We have reported the discovery of several narrow absorption lines
consistent with \fetfive\ or \fetsix\ \ka\ or \kb\ transitions in the
persistent emission spectra of the two dipping LMXBs \nineteen\ and
\twelve. There is no evidence for orbital dependence of the properties
of the lines during the persistent emission of \twelve. An \fetfive\
\ka\ absorption feature is also detected during intensity-selected
dipping spectra of \nineteen, as in the persistent spectrum. The
detection of these lines indicate the presence of a highly inonized
plasma within these sources.

Narrow X-ray absorption lines were first detected from the
superluminal-jet sources \gro\
\cite{1655:ueda98apj,1655:yamaoka01pasj} and \grs\
\cite{1915:kotani00apj,1915:lee02apj}.  ASCA observations of \gro\
revealed the presence of absorption features due to Fe\,{\sc xxv} and
Fe\,{\sc xxvi} which did not show any obvious dependence of their
$EW$s on orbital phase.  \gro\ has been observed to undergo deep
absorption dips \cite{kuulkers98apj} consistent with observing the
source at an inclination angle, $i$, of $60\degmark$--$75\degmark$
\cite{frank87aa}. An inclination of 69.5$\pm$0.3\degree\ is
independently attributed to \gro\ from optical observations by
\cite{1655:orosz97apj}. ASCA observations of \grs\ revealed, in
addition, absorption features due to Ca\,{\sc xx}, Ni\,{\sc xxvii} and
Ni\,{\sc xxviii}. A recent {\it Chandra} HETGS observation of this
source revealed absorption edges of Fe, Si, Mg, and S, as well as
resonant absorption features from Fe\,{\sc xxv} and Fe\,{\sc xxvi} and
possibly Ca\,{\sc xx} \cite{1915:lee02apj}. An inclination of
$\sim$70\degree\ is attributed to \grs, assuming that the jets are
perpendicular to the accretion disk \cite{1915:mirabel94nature}.
Until recently, it was possible that these absorption features were
peculiar to superluminal-jet sources and related in some way to the
jet formation mechanism. With the discovery of narrow absorption
features from the LMXBs \gx\ \cite{gx13:ueda01apjl}, \mxb\
\cite{1658:sidoli01aa}, \bigdip\ \cite{1624:parmar02aa}, and now from
\nineteen\ and \twelve, this appears not to be the case. As proposed
by Kotani et al. \cite{1915:kotani00apj}, ionized absorption features
may be common characteristics of disk accreting systems.  However, it
is interesting to note that 3 of the above 4 LMXBs are dipping
sources. These are systems that are viewed from directions close to
the plane of their accretion disks with $i \sim$60--$80\degmark$
\cite{frank87aa}.  This suggests that inclination angle is important
in determining the strength of these absorption features, which
implies that the absorbing material is distributed in a cylindrical,
rather than a spherical geometry, around the compact object. The
azimuthal symmetry is implied by the lack of any evident orbital
dependence of these features, neither during the persistent emission,
nor during the dipping emission.

The conclusions about the X-ray absorption features discovered in
LMXBs may be summarized as follows:

\begin{itemize}
\item The absorption features are caused by highly ionized ions (iron
or other metals).
\item The early hypothesis of their origin being related to superluminal jets is ruled out.
\item Highly ionized plasmas are probably common characteristis of
disk accreting systems.
\item Absorption features are preferably observed in systems viewed at high inclination.
\item The highly ionized absorbing material is probably distributed in a flat cylindrical geometry around the compact object.
\end{itemize}

\end{document}